\documentclass[%
 reprint,
nofootinbib,
 amsmath,amssymb,
 aps,
 prd,
]{revtex4-2}

\usepackage{graphicx}
\usepackage{dcolumn}
\usepackage{bm}
\usepackage[colorlinks=true, allcolors=blue]{hyperref}

\begin{document}

\preprint{APS/123-QED}

\title{Distinguishing the origin of cosmic birefringence: dark energy, dark matter, and neutrino asymmetry}

\author{Lu Yin}
\affiliation{Department of Physics, Shanghai University, Shanghai, 200444,  China}
\affiliation{Asia Pacific Center for Theoretical Physics, Pohang, 37673, Korea}

\author{Eiichiro Komatsu}
\affiliation{
 Max Planck Institute for Astrophysics, Karl-Schwarzschild-Str. 1, D-85748 Garching, Germany}%
\affiliation{
Ludwig-Maximilians-Universitat M\"unchen, Schellingstr. 4, 80799 M\"unchen, Germany}%
\affiliation{
Kavli Institute for the Physics and Mathematics of the Universe (Kavli IPMU, WPI), UTIAS, The University of Tokyo, Chiba, 277-8583, Japan}

\date{\today}

\begin{abstract}

Cosmic birefringence, the isotropic rotation of the plane of linear polarization of photons from sources at cosmological distances, can be generated by different parity-violating
mechanisms. While the net rotation angle of a single polarized source carries little information about its physical origin, sources at various redshifts $z$ can be used to distinguish between different origins. Parity-sensitive correlations of the $E$- and $B$-mode polarization fields of the cosmic microwave background measured on large and small angular scales probe sources at $z\simeq 10$ and $1100$, respectively. Thus, the detailed shape of the $EB$ power spectrum is sensitive to the
redshift evolution of the birefringence source. In this paper, we calculate the $EB$ power spectrum due to 
a neutrino-asymmetry current, and compare it to axion dark energy and dark matter using a common Chern--Simons framework. We find that the $EB$ power spectrum due to a neutrino asymmetry is suppressed at low multipoles, $\ell\lesssim 20$, similar to the dark matter case. This can be clearly distinguished from the dark energy case, which has no such suppression.
Large-scale polarization measurements, including \textit{LiteBIRD}, can therefore distinguish late-time dark energy birefringence from that
generated at higher redshifts by axion dark matter or neutrino asymmetry. We also find that the evolution of the neutrino density during recombination alters the shape of the $EB$ power spectrum at high multipoles. This signature can be used to distinguish between the neutrino and axion dark matter interpretations, using ground-based experiments such as the Simons Observatory.

\end{abstract}

\maketitle

\section{\label{sec:Introduction}Introduction}
The polarized light of the cosmic microwave background (CMB) is sensitive to new physics that violates
parity symmetry~\cite{Lue:1998mq}, such as ``cosmic birefringence,'' which refers to the uniform rotation of the linear polarization plane due to parity-violating extensions of electromagnetism~\cite{Carroll:1989vb}. Specifically, the correlations among the temperature ($T$) and the $E$- and $B$-mode polarization fields~\cite{Zaldarriaga:1996xe,Kamionkowski:1996ks} contain four parity-even combinations ($TT$, $TE$, $EE$ and $BB$) and two parity-odd combinations ($TB$ and $EB$)~\cite{Lue:1998mq,Feng:2004mq}.

Recent analyses of the $EB$ correlations of polarization data from \textit{Planck} \cite{Planck:2016soo,Minami:2020odp,Diego-Palazuelos:2022dsq,Eskilt:2022wav,Sullivan:2025bva,Remazeilles:2025wzd,Lonappan:2026unb}, \textit{Planck}+\textit{WMAP} \cite{Eskilt:2022cff,Cosmoglobe:2023pgf}, the Atacama Cosmology Telescope (ACT) \cite{AtacamaCosmologyTelescope:2025blo, Diego-Palazuelos:2025dmh, Yin:2026gss}, and ACT+\textit{Planck}~\cite{Eskilt:2026imm} all hint at an isotropic rotation angle of $\beta \simeq 0.3^{\circ}$. If the signal is confirmed to be of cosmological origin, it would imply the discovery of new physics beyond the standard model of elementary particles and fields~\cite{Nakai:2023zdr}, with profound implications for the physical nature of dark matter, dark energy, and quantum gravity (see Ref.~\cite{Komatsu:2022nvu} for a review). However, how can we distinguish between the various origins of cosmic birefringence?

The four-dimensional Chern‑Simons‑like interaction provides a useful effective description of parity-violating photon propagation
\begin{equation}
{\cal L}_{\rm CS}
= \frac12 g_{\rm cs}P_\mu A_\nu \tilde F^{\mu\nu},
\label{eq:cs_intro}
\end{equation}
where $A_\mu$ is the electromagnetic gauge field, 
$\tilde F^{\mu\nu}$ is the dual of the field strength tensor, and $g_{\rm cs}$ is the Chern‑Simons coupling constant. 
The background four‑vector $P_\mu$ encodes the physical origin of cosmic birefringence, which can arise from diverse sources. 

The U(1) gauge invariance of electromagnetism requires that the condition
\begin{equation}
\label{eq:condition}
    \left(\partial_\mu P_\nu-\partial_\nu P_\mu\right)\tilde F^{\mu\nu}=0,
\end{equation}
be satisfied~\cite{Carroll:1989vb}. The simplest possibility is that $P_\mu$ is a time-independent four-vector. 
The constant vector selects a preferred spacetime direction and breaks Lorentz invariance. A purely time-like background preserves spatial isotropy but distinguishes left- and right-handed photon polarization states, producing a uniform rotation of the plane of linear polarization~\cite{Carroll:1989vb}.

A more dynamical realization arises when the background vector is a pseudoscalar‑field gradient,
$P_\mu=\partial_\mu\phi$~\cite{Carroll:1991zs}.
Then, Eq.~\eqref{eq:cs_intro} is $\frac12 g_{\rm cs}
(\partial_\mu\phi)A_\nu\tilde F^{\mu\nu}$ with the mass dimensions given by
$[g_{\rm cs}]={-1}$ and
$[\partial_\mu\phi]=2$,
where the brackets denote mass dimensions in natural units.
Integration by parts yields ${\cal L}_{\rm CS}=-\frac14g_\mathrm{cs}\phi F_{\mu\nu}\tilde F^{\mu\nu}$ up to a boundary term.
This structure was found in early studies of nonmetric electromagnetic couplings and equivalence-principle tests \cite{Ni:1977zz} and axion electrodynamics \cite{Sikivie:1983ip, Wilczek:1987mv, Turner:1987bw}.\footnote{This term also describes the decay of a neutral pion into two photons~\cite{Adler:1969gk,Bell:1969ts}.}
This case naturally describes cosmic birefringence sourced by axionlike late-time dark
energy~\cite{Carroll:1998zi}, dark matter~\cite{Finelli:2008jv,Fedderke:2019ajk,Galaverni:2023zhv}, early dark energy~\cite{Fujita:2020ecn,Murai:2022zur,Eskilt:2023nxm,Yin:2023srb,Kochappan:2024jyf,Yin:2025fmj,Zhang:2026fzj}, and topological defects~\cite{Agrawal:2019lkr,Jain:2021shf,Takahashi:2020tqv,Kaloper:2026slg}. Cosmic birefringence due to a 3-form field, $C_{\mu\nu\rho}$, can also be expressed similarly by noting that $C_{0ij}=0$ and $C_{ijk}=a^3(t)\chi(t)\epsilon_{ijk}$ with $\chi\propto g_\mathrm{cs}\partial_0\phi$ in a homogeneous and isotropic spacetime~\cite{Kamada:2026vbw}. Here, $\epsilon_{ijk}$ is a Levi-Civita symbol.

In this paper, we consider the possibility that the background vector is a matter current~\cite{Geng:2007va,Zhou:2023aqz},
$P_\mu=J_\mu$,
where $J_\mu$ is a fermion-number current. Then, the mass dimensions in natural units are given by 
$[g_{\rm cs}]={-2}$ and
$[J_\mu]=3$.
Unlike in the previous case,
a generic conserved current does not automatically satisfy Eq.~\eqref{eq:condition}.  
To restore the U(1) gauge invariance, one could add a term to the Lagrangian density,
\begin{equation}
{\cal L}_{\rm CS}
\to \frac12 g_{\rm cs}J_\mu \left(A_\nu \tilde F^{\mu\nu}+\partial_\nu S^{\mu\nu}\right),
\label{eq:geng}
\end{equation}
where $S^{\mu\nu}$ is an antisymmetric tensor that transforms in a way that cancels the term induced by the gauge transformation, $A_\mu\to A_\mu+\partial_\mu f$~\cite{Geng:2007va,Ho:2010aq}.

The current acts as an effective timelike Chern-Simons background and rotates the CMB polarization plane.
Since an axion background ($P_\mu=\partial_\mu\phi$) and a neutrino asymmetry ($P_\mu=J_\mu$) can produce the same net polarization
rotation, the rotation amplitude alone does not distinguish their
physical origins. 
In this paper, we show that the discriminating power comes from the shape of the CMB $EB$ power
spectrum across all multipoles.

\section{\label{sec:Model}Model}
We compare three sources of cosmic birefringence: late-time axion dark energy, axion dark matter, and neutrino asymmetry. 
The first two correspond to a pseudoscalar background, $P_\mu=\partial_\mu\phi$, while the neutrino case corresponds to a fermion current, $P_\mu=J_\mu$. Each source generates a time-dependent polarization rotation $\beta(t)$ and produces a CMB $EB$ spectrum.

\textit{Axion Dark Energy:}
We consider a pseudoscalar field with a quadratic potential,
$V(\phi)=m_\phi^2\phi^2/2$.
The homogeneous field obeys
$\ddot\phi+3H\dot\phi+m_\phi^2\phi=0$, 
where $H=\dot a/a$ denotes the Hubble parameter, $a(t)$ is the cosmological scale factor, and an overdot represents differentiation with respect to cosmic time $t$.
The range $m_\phi\lesssim10^{-33}\,{\rm eV}$ corresponds to a dark energy pseudoscalar, whose evolution begins only in the recent universe.
The rotation accumulated after time $t$ is given by~\cite{Carroll:1991zs,Harari:1992ea}
\begin{equation}
\beta_{\rm axion}(t)
=
\frac{g_{\rm cs}}{2}
\left(\phi(t_0)-\phi(t)\right),
\label{eq:alpha_de}
\end{equation}
where $t_0$ denotes the present. In the slowly rolling regime, $\beta_\mathrm{axion}(t)$ admits a universal profile, as discussed in Ref.~\cite{Naokawa:2025shr}. In this regime, the $EB$ power spectrum is proportional to the difference between the $E$- and $B$-mode power spectra, i.e., $C_\ell^{EB}=\frac12\tan\left[4\beta_\mathrm{axion}(t_\mathrm{LSS})\right]\left(C_\ell^{EE}-C_\ell^{BB}\right)$, where ``LSS'' denotes the last scattering surface. Here, $C_\ell^{EE}$ and $C_\ell^{BB}$ also include the effect of cosmic birefringence.

We adopt the benchmark mass of 
$m_{\phi}=10^{-33}\,{\rm eV}$ and take the 
initial field value of $\phi_{\rm ini} = 3.06M_\mathrm{Pl}$ where $M_\mathrm{Pl}$ is the reduced Planck mass, such that its present-day density fraction becomes $\Omega_\mathrm{axion\,DE} = 0.6855$, matching the dark energy abundance of the fiducial flat $\Lambda$ cold dark matter ($\Lambda$CDM) cosmology.

\textit{Axion dark matter:} 
Pseudoscalars with $m_\phi\gtrsim10^{-31}\,{\rm eV}$ behave as a dark matter component. Initially frozen by Hubble friction, the field begins coherent oscillations when its Compton wavelength becomes comparable to the Hubble length. 
After that, its oscillation-averaged energy density redshifts as cold matter~\cite{Hu:2000ke, Marsh:2015xka}. We therefore choose $m_\phi=10^{-30},10^{-29},10^{-28}\,{\rm eV}$ as axion dark matter in this work. 

For the energy component, the total matter abundance ($\Omega_{\rm m}$) is made up of axion dark matter ($\Omega_{\rm axion\,DM}$), cold dark matter ($\Omega_{\rm cdm}$), and baryons ($\Omega_{\rm b}$), written as
$\Omega_{\rm m} = \Omega_{\rm axion\,DM}+\Omega_{\rm cdm}+ \Omega_{\rm b}$,
where $\Omega_{\rm axion\,DM}=10^{-4}$, which is below the most stringent upper limit among the three axion masses ~\cite{Hlozek:2017zzf, Fung:2021fcj, Rogers:2023ezo}. This setup thus provides a conservative benchmark that can be consistently applied when comparing the three mass cases.

Through the same Chern-Simons photon coupling, this axion dark matter
component also rotates the plane of CMB polarization and generates cosmic
birefringence.
The rotation angle is the same as Eq.~\eqref{eq:alpha_de}.
Unlike axion dark energy, whose rotation continues to grow through reionization, axion dark matter completes its coherent oscillations and its rotation angle freezes well before reionization, with the freeze-out redshift set by $m_\phi$. This suppresses $C_\ell^{EB}$ at $\ell\lesssim 20$~\cite{Sherwin:2021vgb}. If the axion mass is on the order of $10^{-28}$~eV, the field evolves during recombination, which alters the shape of $C_\ell^{EB}$ at high multipoles~\cite{Nakatsuka:2022epj}.

\textit{Neutrino asymmetry:}
Finally, we consider birefringence sourced by a 
neutrino asymmetry. Following Ref.~\cite{Zhou:2023aqz}, the fermionic current is coupled to the photon field via
\begin{equation}
{\cal L}_{J} = -\frac12 \frac{b_J}{M^2}
J_\mu A_\nu\tilde F^{\mu\nu},
\label{eq:geng_lagrangian}
\end{equation}
where $M$ is the cutoff scale and $b_J$ is a dimensionless constant.
In a homogeneous and isotropic background, we take
$J_\mu^{\nu_e}=(\Delta n_{\nu_e},\mathbf 0)$,
where $\Delta n_{\nu_e}\equiv n_{\nu_e}-n_{\bar\nu_e}$
is the electron-neutrino number density asymmetry. 
We parameterize the asymmetry by the degeneracy parameter
$\xi_{\nu_e}=\mu_{\nu_e}/T_\nu$, where $\mu_{\nu_e}$ and $T_\nu$ are the neutrino chemical potential and temperature, respectively.

At high redshift, the neutrino acts as a massless particle. For a relativistic Fermi-Dirac distribution, the neutrino asymmetry is~\cite{Kang:1991xa}
\begin{equation}
\eta_{\nu_e}
\equiv
\frac{\Delta n_{\nu_e}}{n_\gamma}
=
\frac{1}{12\zeta(3)}
\left(\frac{T_\nu}{T_\gamma}\right)^3
\left(\pi^2\xi_{\nu_e}+\xi_{\nu_e}^3\right) \simeq 0.249\,\xi_{\nu_e},
\label{eq:eta_nue}
\end{equation}
where $T_\nu/T_\gamma$ is equal to $(4/11)^{1/3}$. By using $n_\gamma(z) = {2\zeta(3)} T_{\gamma0}^3(1+z)^3/{\pi^2}$, we can finally obtain 
\begin{equation}
\Delta n_{\nu_e}(z) \simeq 0.061\,\xi_{\nu_e}T_{\gamma0}^3(1+z)^3 ,
\label{eq:delta_nue}
\end{equation}
where $T_{\gamma0}=2.725$~K is the CMB temperature~\cite{Fixsen:2009ug}.
Thus the neutrino-current source is weighted toward high redshifts.

In the WKB approximation, the modified photon wave equation yields the accumulated rotation angle of~\cite{Zhou:2023aqz}
\begin{equation}
\beta_{\nu_e} =
\frac12\frac{b_J}{M^2}
\int \Delta n_{\nu_e}\,dt
\simeq 0.03  g_{\nu_e} {T_{\gamma 0}^3} \int_0^{z_\mathrm{LSS}} \frac{(1+z)^2}{H(z)} d z,
\label{eq:alpha_nu}
\end{equation}
where
$g_{\nu_e}\equiv b_J\xi_{\nu_e}/{M^2}$
is the effective current-induced Chern-Simons coupling. In the
tomographic $EB$ calculation below, the upper limit $z_\mathrm{LSS}$ is
replaced by the general source redshift $z$, retaining the full
time dependence of $\Delta n_{\nu_e}(z)$.

The three time-dependent rotations are inserted into the 
modified Boltzmann equation as~\cite{Liu:2006uh,Gubitosi:2009eu}
\begin{equation}
\begin{aligned}
{ }_{ \pm 2} \Delta_P^{\prime}+i q \mu_{ \pm 2} \Delta_P
={}&
\tau^{\prime}\left[-{ }_{ \pm 2} \Delta_P+\sqrt{\frac{6 \pi}{5}}{ }_{ \pm 2} Y_2^0(\mu) \Pi(q,\eta)\right]
\\
&\pm 2 i \beta_{ \pm 2}^{\prime} \Delta_P,
\end{aligned}
\end{equation}
where $\eta$ is the conformal time, $q$ the comoving wavenumber, respectively, $\tau(\eta)$ the Thomson scattering optical depth, and
$\Pi(q,\eta)$ the usual polarization source term.
Primes denote derivatives with respect to $\eta$.

Expanding ${ }_{ \pm 2} \Delta_P(\eta,q,\mu)$ using spin-2 spherical harmonics and defining the $E$- and $B$-mode multipole coefficients as ${ }_{ \pm 2} \Delta_{P,\ell}=-(\Delta_{E,\ell}\pm i\Delta_{B,\ell})$, we find the
CMB polarization line-of-sight solution.
Cosmic birefringence introduces a phase
$e^{\pm 2i\beta}$ into the line-of-sight solution for $Q\pm iU$, yielding
\begin{equation}
\begin{aligned}
\Delta_{E,\ell}(q)\pm i\Delta_{B,\ell}(q)
={}&
\frac{3}{4}\sqrt{\frac{(\ell+2)!}{(\ell-2)!}}
\\
&\times
\int_0^{\eta_0} d\eta\,
\tau'e^{-\tau(\eta)}
\Pi(q,\eta)
\frac{j_\ell(x)}{x^2}
e^{\pm 2i\beta} ,
\end{aligned}
\label{eq:eb_transfer}
\end{equation}
where $x=q(\eta_0-\eta)$.
The resulting $EB$ spectrum is
$C_\ell^{EB} = 4\pi\int d\ln q\,
 {\cal P}_{\cal R}(q)
 \Delta_{E,\ell}(q)\Delta_{B,\ell}(q)$,
where ${\cal P}_{\cal R}(q)$ is the primordial curvature power spectrum. 
Finally, we account for the gravitational lensing effect on $C_\ell^{EB}$ according to Ref.~\cite{Naokawa:2023upt}.

\section{\label{sec:Results}Results}

We now compare the birefringence histories and the resulting CMB $EB$ spectra for the three origins introduced above. The axion dark energy and neutrino asymmetry models are computed with a modified version of \texttt{CLASS}~\cite{Lesgourgues:2011re, Blas:2011rf}, while the axion dark matter is computed with \texttt{AxiCLASS}~\cite{Poulin:2018cxd, Poulin:2018dzj}. For each scenario, we numerically solve the background and Boltzmann
equations and incorporate the resulting time evolution of
$\beta$ into the CMB polarization spectra.

\begin{figure}[b]
\includegraphics{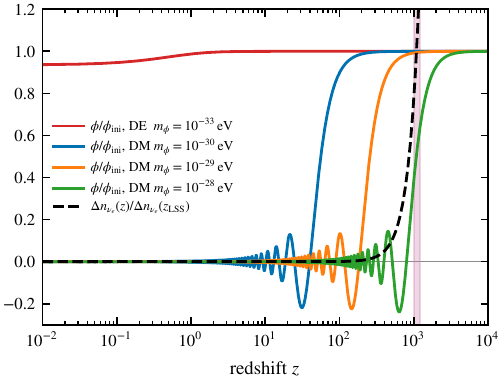}
\caption{\label{fig:evolution} Normalized redshift evolution of the birefringence sources. The solid curves show $\phi/\phi_{\rm ini}$ for the axion dark energy with $m_\phi=10^{-33}\,{\rm eV}$ and the axion dark matter with
$m_\phi=10^{-30},10^{-29},10^{-28}\,{\rm eV}$. A larger $m_\phi$ shifts
the axion oscillation epoch to a higher redshift, bringing the
birefringence source closer to the epoch of recombination (vertical shaded area).
The dashed curve shows the normalized neutrino asymmetry source
$\Delta n_{\nu_e}(z)/\Delta n_{\nu_e}(z_\mathrm{LSS})$. Its evolution differs clearly from that of axion dark matter during recombination.
}
\end{figure}

\begin{figure}[b]
    \centering

    \begin{minipage}{\columnwidth}
        \centering
\includegraphics{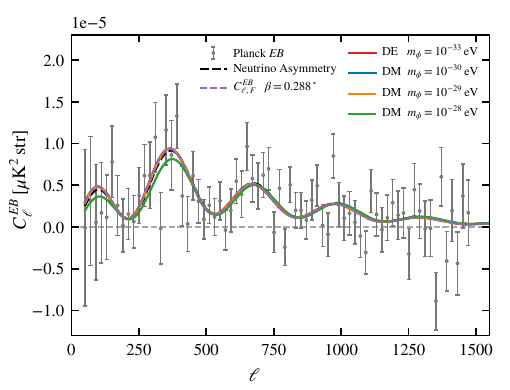}
       \par\vspace{-2mm}
    \end{minipage}

    \vspace{1mm}

    \begin{minipage}{\columnwidth}
        \centering
\includegraphics[width=\linewidth]{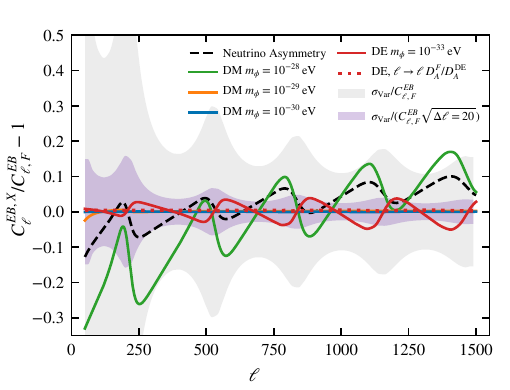}
        \par\vspace{-2mm}
    \end{minipage}
\caption{\label{fig:eb}CMB $EB$ power spectra, $C_\ell^{EB}$, due to neutrino asymmetry (black dashed), axion dark energy (red), axion dark matter (blue, orange, and green), and the constant-rotation formula, $C_{\ell,F}^{EB}$ [Eq.~\eqref{eq:constant}] (purple dashed).  The spectra are normalized to satisfy the \textit{Planck} $EB$ data at $\ell\ge 51$ (points with error bars shown in the top panel)~\cite{Eskilt:2023nxm}. The bottom panel shows the ratio of each scenario to $C_{\ell,F}^{EB}$. The gray and purple shaded areas show  cosmic variance error bars for the dark energy case and those divided by $\sqrt{20}$. The neutrino asymmetry scenario can be distinguished from all other scenarios with high statistical significance.}
\end{figure}

For the axion dark matter benchmarks, we take
$m_\phi=10^{-30},10^{-29},10^{-28}\,{\rm eV}$. For each mass, the
initial field value is obtained by shooting for the target present-day
axion abundance, $\Omega_{\rm axion\, DM}=10^{-4}$.
This makes the axion a subdominant component of the total matter
density, with the remaining abundance supplied by cold dark
matter and baryons. With this choice, the background evolution remains close to the fiducial flat $\Lambda$CDM cosmology and satisfies the existing constraints. This allows us to focus on the birefringence spectral shape.

Figure~\ref{fig:evolution} displays the time evolution of the
normalized cosmic birefringence sources. The axion dark energy field remains
frozen through reionization and
begins to evolve only at late times, around $z\simeq1$, thereby leaving a characteristic imprint on the $EB$ spectrum.
In contrast, the axion dark matter benchmarks, $m_\phi=10^{-30},10^{-29},10^{-28}\,{\rm eV}$, begin to oscillate when $H\simeq m_\phi$, with the onset shifting to a higher redshift for a larger mass. In these \texttt{AxiCLASS} runs, the fluid approximation is switched
off, so the plotted evolution explicitly resolves the oscillations of $\phi$ rather than using an oscillation-averaged effective fluid description. The neutrino asymmetry density evolves as ${\Delta n_{\nu_e}(z)}/{\Delta n_{\nu_e}(z_{\rm LSS})} =({1+z})^3/({1+z_\mathrm{LSS}})^3$, as shown by the black dashed curve.
Since its evolution differs clearly from that of axion dark matter during recombination, we also expect the shape of $C_{\ell,F}^{EB}$ due to neutrino asymmetry to differ.

Figure~\ref{fig:eb} shows the $EB$ spectra at high multipoles, $\ell\ge 51$, induced by the three
birefringence sources. Each coupling constant is obtained from a likelihood function similar to Eq.~(8) in Ref.~\cite{Eskilt:2023nxm}, but with a Gaussian prior for the miscalibration angle, $\alpha$, and a standard deviation of $\sigma_\alpha=0.11^\circ$, taken from the calibration uncertainty for the 353~GHz channel~\cite{Diego-Palazuelos:2022dsq}.
The top panel shows that the \textit{Planck} data are still too noisy to distinguish among the three scenarios. For reference, the purple dashed curve shows a constant-rotation formula constructed from the lensed $EE$ and $BB$ spectra of $\Lambda$CDM,
\begin{equation}
C_{\ell,F}^{EB}=
\frac{1}{2}\sin(4\beta)
\left(C_\ell^{EE,\Lambda\mathrm{CDM}}-C_\ell^{BB,\Lambda\mathrm{CDM}}\right),
\label{eq:constant}
\end{equation}
with $\beta=0.288^\circ$.

The bottom panel shows the ratio of each scenario to $C_{\ell,F}^{EB}$ along with cosmic variance limited error bars,  
$\sigma\!\left(C_\ell^{EB}\right) = \sqrt{{\rm Var}\!\left(C_\ell^{EB}\right)}$,
where ${\rm Var}\!\left(C_\ell^{EB}\right) = \left[ C_\ell^{EE}C_\ell^{BB} + \left(C_\ell^{EB}\right)^2\right]/({ 2\ell+1})$~\cite{Knox:1995dq}, for the dark energy case. The purple shaded area shows the error bars when binned with $\Delta\ell=20$, ${\rm Var}\!\left(C_\ell^{EB}\right)/20$. The $\chi^2$ analysis shows that the neutrino asymmetry scenario can be distinguished from all other scenarios with high statistical significance. For example, neutrino asymmetry can be distinguished from $C_{\ell,F}^{EB}$ and axion dark matter of $m=10^{-28}$~eV with $\Delta\chi^2\simeq 170$ and 180, respectively, for $\ell=51-1490$. We find similar results for other cases. The statistical significance increases as the maximum multipole, $\ell_\mathrm{max}$, increases. 

What is the origin of the differences in the shape of $C_\ell^{EB}$ at high multipoles? As shown in Fig.~\ref{fig:evolution} and explained in Ref.~\cite{Nakatsuka:2022epj}, the difference at high multipoles is due to the difference in the evolution of the cosmic birefringence source during recombination. The evolution of the neutrino asymmetry differs clearly from that of the others. The evolution of axion dark matter with $m=10^{-28}$~eV also differs from other axion scenarios. However, those for $m=10^{-29}$ and $10^{-30}$~eV are the same during recombination; thus, they are indistinguishable. In contrast, axion dark energy with $m=10^{-33}$~eV appears to produce a different shape, as it slightly changes the angular diameter distance to the last scattering surface compared to the $\Lambda$CDM value. This does not occur in dark matter cases because the energy density of axion dark matter is negligibly small.

Using specifications of the Simons Observatory's Large Aperture Telescope~\cite{SimonsObservatory:2018koc} that were used in the analysis of Ref.~\cite{Nakatsuka:2022epj}, we find that the neutrino asymmetry can be distinguished from $C_{\ell,F}^{EB}$ and the axion dark matter of $m=10^{-28}$~eV with $\Delta\chi^2\simeq 15$ and 29, respectively, for $\ell=100-2500$. The other cases fall between these two values. Incorporating the uncertainty in the miscalibration angle with $\sigma_\alpha=0.11$~deg~\cite{AtacamaCosmologyTelescope:2025blo,Diego-Palazuelos:2025dmh}, we find $\Delta\chi^2\simeq 10$ and 25, respectively.

\begin{figure}
\includegraphics{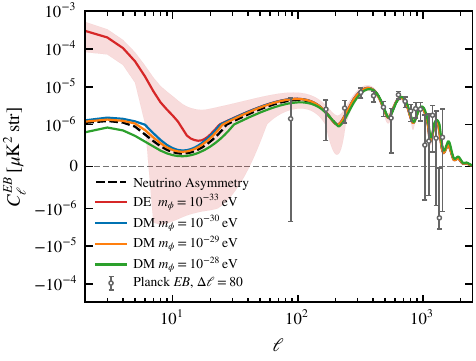}
\caption{\label{fig:lowell}CMB $EB$ power spectra including lower multipoles. The axion dark energy benchmark shows
an enhanced $EB$ signal that can be clearly distinguished from that of the neutrino asymmetry scenario. The \textit{Planck} data points are rebinned with $\Delta\ell=80$, whereas the original bin size shown in the top panel of Fig.~\ref{fig:eb} is $\Delta\ell=20$.}
\end{figure}

Moving to low multipoles, Figure~\ref{fig:lowell} expands the multipole range to $\ell=2$ on a logarithmic
 scale. For clarity, the \textit{Planck} $EB$ bandpowers are rebinned with a bin size of $\Delta\ell=80$ using inverse-variance weighting.  
Among the benchmarks considered, the $m_\phi=10^{-33}\,{\rm eV}$ axion dark energy produces the largest low-$\ell$ power and a pronounced upward bump on large scales, since it sources the rotation at late times. 
For the dark matter benchmarks, the low-$\ell$ amplitude decreases as $m_\phi$ increases from $10^{-30}$ to $10^{-28}\,{\rm eV}$. As shown in Fig.~\ref{fig:evolution} and explained in Ref.~\cite{Nakatsuka:2022epj}, a heavier field begins oscillating at a higher redshift, allowing its amplitude to decay more strongly before reionization. Consequently, its value during reionization is closer to the present-day value, resulting in less rotation of the reionization polarization and hence a weaker low-$\ell$ $EB$ signal. Similarly, since the neutrino asymmetry density evolves as $a^{-3}$, cosmic birefringence from the reionization epoch is small and the low-$\ell$ $EB$ power spectrum is suppressed. 

A detailed statistical analysis~\cite{Carralot:2026kps} shows that cosmic variance-limited $EB$ data on large angular scales can distinguish between dark energy and dark matter cases with high significance. Therefore, such data can also distinguish between dark energy and neutrino asymmetry cases considered in this paper.
Although current measurements at these multipoles are noisy, forthcoming large-scale polarization experiments, including \textit{LiteBIRD}~\cite{LiteBIRD:2025yfb}, will offer substantially improved sensitivity to this signature.

For cosmic birefringence due to neutrino asymmetry, \textit{Planck} $EB$ data constrain the effective coupling, giving 
${M}/{\sqrt{b_J\xi_{\nu_e}}} = 44.32^{+12.32}_{-6.74}\,{\rm TeV}$  
at $68\%$ credibility, after marginalizing over $\alpha$ with $\sigma_\alpha=0.11^\circ$~\cite{Diego-Palazuelos:2022dsq}. The data contribution to the
chi-square at the best-fit point is $\chi^2=65.8$. The fit uses 72 \textit{Planck} $EB$ bandpowers and two free parameters, corresponding to 70 degrees of freedom. This value is consistent with the constraint shown in Ref.~\cite{Zhou:2023aqz}. However, extracting a physical scale, $M$, requires knowing the dimensionless coupling $b_J$ and the neutrino degeneracy parameter $\xi_{\nu_e}$. We take $\xi_{\nu_e}=0.05^{+0.03}_{-0.02}$ from the EMPRESS
measurement~\cite{Matsumoto:2022tlr} and adopt $b_J=0.01$, i.e., the same order of magnitude as the fine structure constant. We obtain
\begin{equation}
M = 991.1^{+275.4}_{-150.8}\,{\rm GeV}
\left(\frac{b_J}{0.01}\right)^{1/2}
\left(\frac{\xi_{\nu_e}}{0.05}\right)^{1/2}.
\label{eq:M_constraint}
\end{equation}
Therefore, this fit yields $M\simeq 1$~TeV, indicating that, if the \textit{Planck} $EB$ birefringence signal is interpreted as a neutrino current effect, then the corresponding effective interaction is associated with a particle physics scale close to the electroweak scale. Existing estimates from rare
$Z$-decay data only place weak limits on this operator and do not
exclude this parameter range~\cite{Zhou:2023aqz}.

\section{\label{sec:Conclusions}Conclusions}

As demonstrated in Refs.~\cite{Nakatsuka:2022epj,Murai:2022zur,Galaverni:2023zhv,Carralot:2026kps},
the physical origin of cosmic birefringence, such as axion dark energy and dark matter, can be
distinguished through the shape of the CMB $EB$ power spectrum. In this paper, we have extended the analysis to include cosmic birefringence due to neutrino asymmetry as proposed in Refs.~\cite{Geng:2007va,Zhou:2023aqz}. Using a
single Chern-Simons framework given in Eq.~\eqref{eq:cs_intro}, we compared three sources normalized to the same \textit{Planck} $EB$ power spectrum data at high multipoles (Fig.~\ref{fig:eb}), which include axion dark energy, axion dark matter, and a neutrino asymmetry current. 
Although these sources can produce the same net rotation angle $\beta$, they have different redshift evolution histories. We found that this difference is imprinted across all multipoles.

For the benchmark masses considered here, the neutrino asymmetry scenario can be distinguished from the others across all multipoles. The detailed shape of the high-$\ell$ $EB$ power spectrum can distinguish between the neutrino asymmetry and axion dark matter scenarios. The suppression at low $\ell$ can distinguish between the neutrino asymmetry and axion dark energy scenarios. All of this can be understood simply from the evolution of the cosmic birefringence source shown in Fig.~\ref{fig:evolution}. Recently, Ref.~\cite{Co:2026rga} has shown that cosmic birefringence can arise from rotations of the axion field in field space with $\dot\phi\propto a^{-3}$. Since this time dependence is the same as that of the neutrino asymmetry scenario, we also expect their $EB$ spectra to be similar.

For the neutrino-current interpretation, the \textit{Planck} $EB$
data give
$M\simeq 1$~TeV [Eq.~\eqref{eq:M_constraint}].
This connects the CMB birefringence signal to a TeV-scale effective interaction in the neutrino sector, while existing rare $Z$-decay bounds do not exclude this parameter range. In conclusion, future large-scale polarization measurements, especially \textit{LiteBIRD} observations of the low-$\ell$ $EB$ signal~\cite{LiteBIRD:2025yfb}, and high-$\ell$ $EB$ measurements by ground-based experiments such as the Simons Observatory~\cite{SimonsObservatory:2018koc} can distinguish cosmic birefringence due to neutrino asymmetry from other sources. Our study has demonstrated that the shape of the $EB$ power spectrum is a powerful test of fundamental physics beyond the standard model of elementary particles and fields.

\begin{acknowledgments}

L. Yin thanks the Max Planck Institute for Astrophysics (MPA) for hospitality, where this work was initiated. She also thanks Prof. C.Q. Geng and members of the Department of Physical Cosmology at the MPA for fruitful discussions. This work was supported in part by the Natural Science Foundation of Shanghai 24ZR1424600, the Deutsche Forschungsgemeinschaft (DFG, German Research Foundation) under Germany's Excellence Strategy - EXC-2094/2 - 390783311, and the European Research Council (ERC) under the European Union's Horizon 2020 research and innovation program (NewPhysCMB, Grant agreement No. 101264428). Views and opinions expressed are however those of the authors only and do not necessarily reflect those of the European Union or the ERC Executive Agency. Neither the European Union nor the granting authority can be held responsible for them.
 
\end{acknowledgments}




\bibliographystyle{apsrev4-2}
\bibliography{apssamp}

\end{document}